\documentstyle[12pt,oldlfont,titlepage,a4,psfig]{article}

\begin{document}
\begin{center}
{\LARGE \bf Aging and memory phenomena in magnetic and 
transport properties of vortex matter: a brief review}

\vspace{2cm}

{\large \sl Mario Nicodemi$^{1,2}$ and Henrik Jeldtoft Jensen$^{1}$}
\vspace{2cm}
 
{\em $^{1}$Department of Mathematics, Imperial College, 180 Queen's Gate, \\
London SW7 2BZ, UK \\ $^{2}$Universit\'a di Napoli ``Federico II'',
Dip. Scienze Fisiche, INFM and INFN, Via Cintia, 80126 Napoli, Italy\\
e-mail: m.nicodemi@ic.ac.uk and h.jensen@ic.ac.uk
}

\vspace{2cm}
 
{\bf Abstract}
 
\end{center}

There is mounting experimental evidence that strong off-equilibrium phenomena, 
such as ``memory'' or ``aging'' effects, play a crucial role in the physics of 
vortices in type II superconductors. In the framework of a 
recently introduced schematic vortex model, we describe the out of equilibrium 
properties of vortex matter. 
We develop a unified description of ``memory'' phenomena in magnetic and 
transport properties, such as magnetisation {\em loops} and their 
``anomalous'' {\em $2$nd peak}, logarithmic {\em creep}, 
``anomalous'' {\em finite creep rate} for $T\rightarrow 0$, 
{\em ``memory''} and {\em ``irreversibility''} of I-V characteristics, 
{\em time dependent} critical currents, 
{\em ``rejuvenation''} and {\em ``aging''} of the system response. 

\vspace{4cm}

\pagebreak

\section{Introduction}

The properties of vortex dynamics in type II superconductors crucially affect 
the overall system behaviour and have, thus, relevant effects in technological 
applications \cite{blatter,brandt,yeshurun,cohen}. 
In particular, in the last few years it has been discovered that vortex matter 
exhibits important, even dominant, history dependent phenomena in magnetic and 
transport properties, such as memory and hysteresis in magnetisation 
curves along with irreversibility and aging in I-V characteristics 
(see 
\cite{yeshurun,cohen,Bhatta,solid-solid-yeshurun,Giller,Perkins,Roy,Kokka,Paltiel,Valenzuela,Martinoli,vanderBeek,Papad,Fuchs,higgins,andrei}
and references therein). 
These phenomena are markedly {\em out of equilibrium} effects and, here, 
we discuss their features as they emerge from the off equilibrium 
dynamics of vortex matter 
(see Ref.s in \cite{blatter,glass_rev,vmc,ldcp,fldm}). 

The above experimental findings have interesting analogies with 
``memory'' and ``aging'' 
effects observed in other glass formers, such as polymers, supercooled 
liquids or random magnets \cite{Angell,glass_rev}. 
Interestingly, ``glassy'' dynamics have important universal structural 
properties \cite{Angell,glass_rev}. 
Off equilibrium features arise when typical experimental probing times 
get much shorter compared to the system long (often inaccessibly long) 
intrinsic relaxation time scales. These can become huge 
at low temperatures or high densities, where a true equilibrium glass 
phase transition in some cases can be also found. 
Such glass transitions are called ``ideal'' \cite{Angell} 
because, as just stated, equilibrium might be hardly approached. 
In facts, the notion of ``glassy phases'' 
has been repeatedly used in relation to new {\em equilibrium} phases of 
vortex matter \cite{blatter,mfisher,natter,Nelson,GiLeD}. 
We are concerned here, however, with the general properties of 
{\em off equilibrium dynamics} of vortices, not with their equilibrium 
transitions \cite{nota_notc}. 

We consider a schematic model \cite{nj_1} that contains the essential degrees 
of freedom of a vortex system and is simple enough to allow a complete 
understanding of its off equilibrium dynamics in the same perspective 
successfully used for other glassy systems \cite{Angell,glass_rev}. 
The model (a coarse grained\cite{barford,bassler} system of repulsive particles wandering  
in a pinning landscape in presence of a thermal bath and an external drive) 
describes several phenomena of vortex physics, ranging from 
a {\em reentrant phase diagram} in the $(B,T)$ (field-temperature) plane,
to the anomalous {\em ``second peak''} in magnetisation loops
(the ``fishtail''), {\em logarithmic creep} and {\em ``aging''} 
of magnetic relaxation, the {\em finite creep rates} for $T\rightarrow 0$ 
(without use of ``quantum effects''), 
{\em ``memory''} and {\em history} dependent behaviours in vortex flow 
and in I-V characteristics, and many others \cite{nj_1}.

We describe here the properties of such a model and depict a unified picture 
of creep and transport measurements. In particular, the system dynamics can 
be described by identifying its important time scales and their dependence on
temperature, magnetic field and applied electrical current. 
We also suggest new experiments that will help to clarify the nature 
of glassy aspects in superconductors. 

In the next section we introduce the model \cite{nj_1} and in the following 
we systematically compare its behaviours with experiments on magnetic 
and  transport properties. Finally in the conclusions we give an overview of 
our scenario of off equilibrium phenomena in vortex matter.

\section{The R.O.M. Model}

Vortices in type II superconductors are described by the Ginzburg-Landau 
equations. The typical high vortex densities and long interaction range 
imply that the vortex system is strongly interacting. In brief, this makes 
the theoretical description of its equilibrium and, even worse, dynamical 
properties highly non trivial \cite{blatter,yeshurun}. 

An appealing and much used approximation for the {\em microscopic} vortex 
dynamics is based on Molecular Dynamics (MD) simulations 
(see for instance Ref.s in \cite{henrik,Nori,Gronbech-Jensen,Zimanyi}). 
However, even this simplified approach is hardly feasible to explore the 
physics of the 
long time and space scales, low temperatures and high densities region 
where glassy features substantially appear \cite{Zimanyi}. 
Alternatively it was proposed to use schematic discrete time and space models 
\cite{jensen1} to study vortex properties. 

\begin{figure}
\vspace{-0.5cm}
\hspace{-0.0cm}
\centerline{\psfig{figure=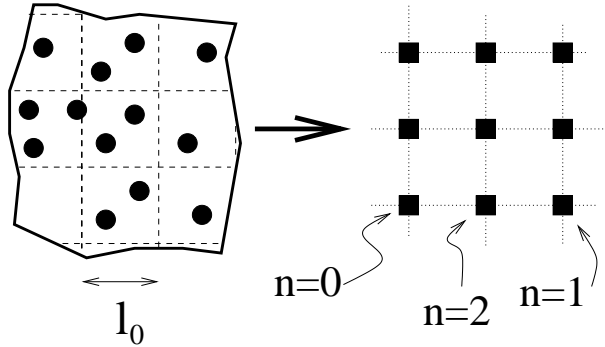,width=8cm,angle=0}}
\vspace{-0.2cm}
\caption{A schematic plot of the procedure introduced to define the ROM 
lattice model. The original vortex system (left), coarse grained 
in ``cells'' of size $l_0$, is mapped into a lattice field model (right).}
\label{lattice}
\end{figure}

More generally, to describe the relevant degrees of freedom of the vortex 
system one can introduce useful {\em coarse graining} methods, successfully 
applied to deal with many other multiscale problems (such as magnetism or 
crystals defects, see Ref.s in \cite{CG}). 
For clarity, let's consider the simple case of a system of straight parallel 
vortex lines, corresponding to a magnetic field $B$ along the $z$-axis, where 
vortices interact through a two-body potential \cite{brandt}: 
\begin{equation}
A(r)= {\phi^2_0\over 2\pi\lambda'^2}
\left[ K_0(r/\lambda')- K_0(r/\xi')\right], 
\label{vortex_pot}
\end{equation}
$K_0$ being the MacDonald function, $\xi$ and $\lambda$ the correlation 
and penetration lengths 
($\xi'=c\xi/\sqrt{2}$, $\lambda ' = c\lambda$, $c=(1-B/B_{c2})^{-1/2}$).
A simple application of the above methods in the present case, proposed 
in \cite{bassler,nj_1}, consists in coarse graining the vortex 
system in the $xy$-plane by introducing a square grid of lattice spacing, 
$l_0$, of the order of the London length, $\lambda$ 
\cite{nota_l0} (see Fig.\ref{lattice}). 

By this procedure, the original vortex system is mapped into a lattice 
model characterised by a classical field, $n_i$, representing the number of 
vortices on the $i$-th coarse grained cell (see Fig.\ref{lattice}). 
The presence in superconductors of an upper critical field, $B_{c2}$, implies 
that $n_i$ must be an integer number smaller than 
$N_{c2}=B_{c2}l_0^2/\phi_0$ \cite{nj_1} 
($\phi_0=hc/2e$ is the flux quantum). 
The Hamiltonian of the coarse grained model is \cite{nj_1}:
\begin{equation}
{\cal H}= \frac{1}{2} \sum_{ij} n_i A_{ij} n_j
-\frac{1}{2} \sum_i A_{ii} |n_i| - \sum_i A^p_i n_i
\label{H}
\end{equation} 
The first two terms of ${\cal H}$ describe the repulsion between the vortices 
and their self energy, and the last the interaction with a random pinning 
background. For sake of simplicity, since $l_0\sim \lambda$, 
we can consider the simplest version of ${\cal H}$:
we choose $A_{ii} = A_0=1$; $A_{ij}=A_1<A_0$ if $i$ and $j$ are nearest
neighbours; $A_{ij}=0$ otherwise; the random pinning is taken to be 
delta-distributed
$P(A^p)=(1-p)\delta(A^p)+p\delta(A^p-A^p_0)$ (see \cite{nota_L}).
We express all energy scales in units of $A_0$ and, in particular, 
consider the important ratio $\kappa^*=A_1/A_0$. 
The existence of two possible orientations of the vortices  
can be taken into account by giving the particles, $n_i$, a ``charge'' 
$s_i=\pm 1$ \cite{blatter,brandt}. 
Neighbouring particles with opposite ``charge'' annihilate. 
The external applied field controls the overall system ``charge density'' 
and thus a chemical potential term $-\mu\sum_i s_in_i$ must be added to the 
Hamiltonian in Eq.(\ref{H}) (where $n_i$ is replaced by $s_in_i$). 

\begin{figure}
\vspace{-2cm}
\hspace{-1.8cm}
\centerline{\psfig{figure=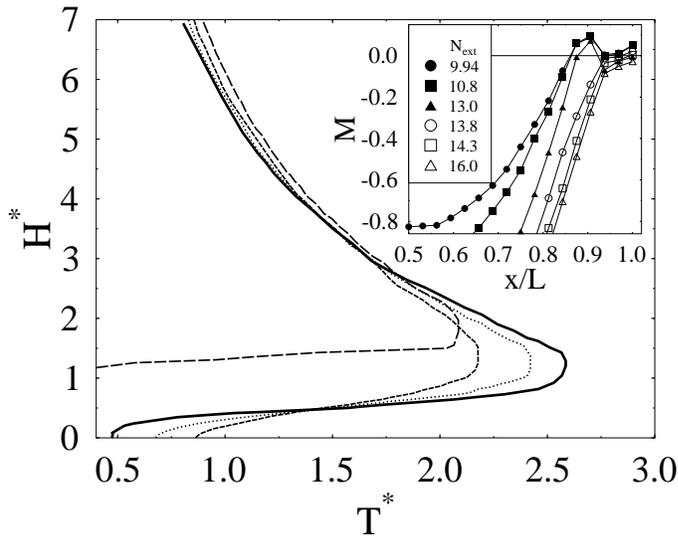,width=8cm,angle=-90}}
\vspace{-1.6cm}
\caption{{\bf Main frame} The mean field phase diagram of the 
ROM model in the plane $(H^*,T^*)$ ($T^*=T/A_1$ and $H^*=\mu/k_BT$ are the 
dimensionless temperature and chemical potential of the applied field), 
for $\kappa^*=10$ and $A^p_0=0.0;0.5;0.75$ (res. full, dotted and dashed 
lines) and $\kappa^*=3.3$ and $A^p_0=0.0$ (long dashed line). 
{\bf Inset} The magnetisation Bean profile, $M(x)$, as a function of the 
transversal spatial coordinate $x/L$ ($L$ is the system linear size), 
(for the shown values $N_{ext}$), in the 2D ROM model 
($T=0.3$, $\gamma=1.1~10^{-3}$). 
Notice the change in shapes when $N_{ext}$ crosses 
$N_{sp}\simeq 13.5$ (filled v.s. empty symbols). 
}
\label{mfpd}
\end{figure}


A standard mean field replica theory \cite{nj_1} 
allows to evaluate the equilibrium phase diagram in the field-temperature 
plane of the above Hamiltonian, as shown in Fig.\ref{mfpd}. 
In absence of disorder it has, at low temperatures, 
a reentrant order-disorder transition in agreement with 
predictions \cite{blatter} and experiments on vortices in superconductors 
(see Ref.\cite{blatter,yeshurun,cohen} or, for instance, 
data on 2H-NbSe$_2$ superconductors from Ref.\cite{Bhatta}). 
For moderate values of the pinning energy ($A^p_0\le A_1$), a second order
transition still takes place, which at sufficiently strong pinning
is expected to become a ``glassy'' transition, as is seen in
Random Field Ising Models \cite{Nattermann}. 
The extension of the low $T$ phase shrinks by increasing $A^p_0$ 
(i.e., the highest critical temperature, $T_m^*$, decreases)
and the higher is $\kappa^*$ the smaller the reentrant region
(facts in agreement with experiments, 
see for instance Ref.s in \cite{yeshurun,cohen,exp2ndpeak,Bhatta}). 
The above phase diagram 
can help to compare experimental and model temperature/field scales. 


We now go beyond mean field theory and discuss the dynamics of the 
model. First we consider the case where the external current 
is absent, i.e., there is no Lorentz drive on vortices. 
The simplest consistent approach to simulate the system relaxation at 
non-zero temperatures is a Monte Carlo Kawasaki dynamics \cite{Binder}
on a square lattice of size $L$ at a temperature $T$ (see \cite{nota_L}).
This is a very standard approach in computer simulations of 
dynamical processes in complex fluids \cite{Binder}. In particular, 
we consider a system periodic in the $y$-direction. Its two edges 
parallel to the $y$-axis are in contact with a vortex reservoir, i.e., an
external magnetic field, of density $N_{ext}$.
Particles can enter and leave the system only through the reservoir.

The above model, called ROM (Restricted Occupancy Model), is described in full
details in \cite{nj_1}. It is extremely schematic, thus, also fully tractable, 
and, interestingly, it is able to describe many of the experimental 
observations on magnetic and transport properties of vortex physics. 

\begin{figure}[htb]
\vspace{-2cm}
\hspace{-1.8cm}
\centerline{\psfig{figure=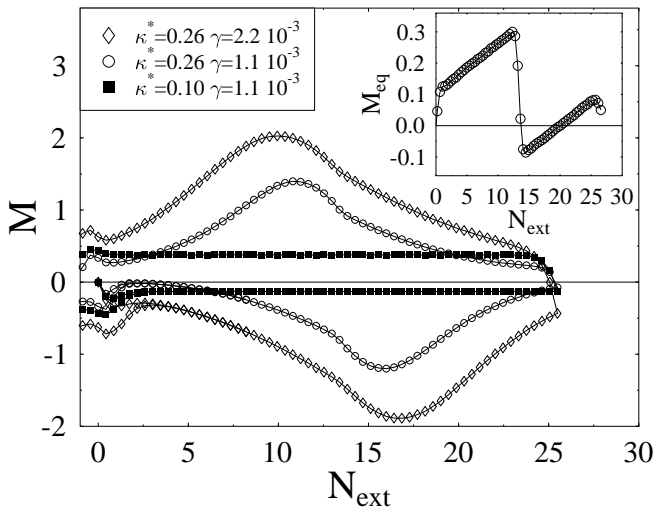,width=8cm,angle=-90}}
\vspace{-1.8cm}
\caption{{\bf Main frame} The magnetisation, $M$, as a function of the applied
field density, $N_{ext}$, in the ROM model at $T=0.3$ for the shown sweep 
rates $\gamma$ and $\kappa^*$. 
Notice the appearance of a ``second magnetisation peak'' when $\kappa^*$ is 
large enough. {\bf Inset}
The equilibrium value of $M$ (i.e., when the field ramp rate 
$\gamma\rightarrow 0$) at $T=0.3$ ($\kappa^*= 0.26$).}
\label{loops}
\end{figure}

\section{The Magnetisation}

The simplest quantity to characterise the vortices system is the 
magnetisation, which we now consider. In this section we will draw a picture 
of time dependent magnetic features such as 
magnetisation {\em loops}, their {\em $2nd$ peak}, {\em ``aging''} creep, 
as well as phenomena like the {\em finite creep rate}, 
$S_a>0$, found when $T\rightarrow 0$. 

The system is prepared by zero field cooling at a given $T$ and 
then increasing the external field, $N_{ext}$, with a constant rate, $\gamma$. 
During the ramp of $N_{ext}$ we record the magnetisation 
\begin{equation}
M(t)=N_{in}(t)-N_{ext}(t) ~ .
\end{equation}
Here $N_{in}=\sum_i n_i/L^d$ and the Monte Carlo time, $t$, is 
measured in units of complete Monte Carlo lattice sweeps. 

\subsection{Magnetisation loops}

At low temperatures pronounced hysteretic magnetisation loops are seen 
when $M$ is plotted as a function of $N_{ext}$ (see Fig.~\ref{loops}). 
Furthermore, when the parameter $\kappa^*=A_1/A_0$ ($\kappa^*$ can be directly 
related to the Ginzburg-Landau parameter $\kappa=\lambda/\xi$ \cite{nj_1}) 
is high enough, a definite {\em second peak} (``fish-tail'') appears in $M$.
Very similar magnetisation data are observed in a number of different
superconductors from intermediate to high $\kappa$ values (see, for instance, 
\cite{yeshurun,cohen,exp2ndpeak,Bhatta,solid-solid,solid-solid-yeshurun,Giller,Perkins,Roy,Kokka,Paltiel,Fuchs,GurevichV} and references therein). 

The actual shape of loops depends on the system parameters (and its size). 
In particular, the sweep rate of the external field, $\gamma$, is very 
important, as shown in Fig.\ref{loops}. As soon as the inverse of the 
sweep rate is smaller than the system characteristic relaxation time 
(see below) strong hysteresis effects are present.
Although the second peak does depend on dynamics through $\gamma$, 
in the ROM model it is related to a new phase 
transition:
in the $\gamma\rightarrow0$ limit, its location, $N_{sp}$, is associated with 
a sharp jump in $M_{eq}\equiv\lim_{\gamma\rightarrow 0} M(\gamma)$ 
(see inset of Fig.\ref{loops}).
These findings are consistent with experiments
(for instance, see Ref.\cite{Bhatta,solid-solid,solid-solid-yeshurun,Giller,Perkins,Roy,Kokka,vanderBeek,cruz,Fuchs,Marchevsky}) and
to some extents reconcile previously proposed opposite descriptions
(``static'' v.s. ``dynamic''). 

\subsection{History dependent relaxation: ``aging'' creep} 

The presence, at low temperature, of sweep rate dependent hysteretic cycles,
slowly relaxing magnetisation, and similar effects, indicate that our system, 
on the observed time scales, is not at equilibrium. 
We turn, therefore, to the theoretical description of the system dynamics by 
investigation of two times correlation functions. 
At the given working value of the applied field, we also record the magnetic 
correlation function, $C(t,t_w)$ (with $t>t_w$), which gives richer 
information than $M(t)$ \cite{nota_MC}: 
\begin{equation}
C(t,t_w)
=\langle [M(t)-M(t_w)]^2\rangle ~ .
\label{cor}
\end{equation}

\begin{figure}[ht]
\vspace{-1.5cm}
\hspace{-2.2cm}
\centerline{\psfig{figure=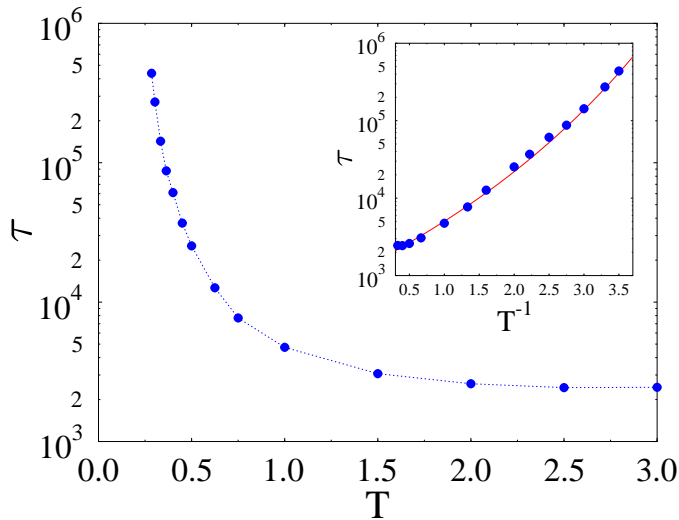,width=8cm,angle=-90}}
\vspace{-2.2cm}
\caption{
{\bf Main frame } The system equilibration time, $\tau$, from 
eq.(\ref{strexp_C}), enormously 
grows by decreasing the temperature $T$ (here $N_{ext}=10$). 
Below the crossover temperature $T_g\sim 0.25$, $\tau$ is larger than 
the observation window.
{\bf Inset } Close to $T_g$,
$\tau$ plotted as a function of $1/T$ approximately shows
a Vogel-Tamman-Fulcher behaviour, the continuous line 
(see eq.(\ref{VTF})).}
\label{CT1par_varie_temp}
\end{figure}

At not too low temperatures, for instance at $T=1.0$ (a 
comparison of such a $T$ value with experimental scales can be derived 
from Fig. \ref{mfpd}), the system creep
is characterised by finite relaxation times and no ``aging'' is seen:
$C(t,t_w)$ is a function of $t-t_w$.
At long times, $C(t,t_w)$ is well fitted by the so called
Kohlrausch-Williams-Watts (KWW) law \cite{nj_1}:
\begin{equation}
C(t,t_w)\simeq C_{\infty}\left\{1-e^{-[(t-t_w)/\tau]^{\beta}}\right\} ~ .
\label{strexp_C}
\end{equation}  
Eq.(\ref{strexp_C}) defines the characteristic time scale of magnetic 
relaxation, $\tau$. This is a crucial quantity to be considered when 
dealing with dynamical aspects of magnetisation. 
The Kohlrausch-exponent, $\beta$, and $\tau$ strongly depend on $T$ 
(a fact to be discussed below, see Fig.\ref{CT1par_varie_temp}) 
and on the applied field $N_{ext}$ \cite{nj_1}.
The pre-asymptotic dynamics (i.e., $t-t_w<<\tau$) is
also interesting and characterised by various regimes. In particular, for not
too short times, a power law is observed over several decades. 

The scenario described for $T=1.0$ is found in a broad
region at low temperatures. However, around $T=0.5$,
a steep increase of $\tau$ is found (see Fig.\ref{CT1par_varie_temp}).
For instance at $N_{ext}=10$, for temperatures below $T_g\simeq 0.25$, the 
characteristic time gets larger than our recording window and the system 
definitely loses contact with equilibrium.
The crossover temperature, $T_g(N_{ext})$ (which may be a function of 
$\gamma$) has a physical meaning similar to the so called phenomenological 
glass transition point in supercooled liquids\cite{Angell}. 
The presence of an underlying ``ideal'' glass transition point, 
$T_c(N_{ext})$, is often located by some fit of the high $T$ data for $\tau$ 
(see inset of Fig.\ref{CT1par_varie_temp}), 
such as a Vogel-Tamman-Fulcher (VTF) (or a power) law: 
\begin{equation}
\tau=\tau_0 \exp\left({E_0\over T-T_c}\right) ~ .
\label{VTF}
\end{equation}    
Our data in 2D, are consistent with $T_c=0$. 
Interestingly, the VTF behaviour found here is in agreement with results 
from Molecular Dynamics simulations of more realistic 
London-Langevin models \cite{nj_1,henrik,Nori,Gronbech-Jensen,Zimanyi}. 
In particular, the analogies with ``window glasses'' have been also outlined 
in the first of Ref.s \cite{Zimanyi}. 
A VTF behaviour has been already experimentally observed in measures 
on samples resistivity (see \cite{sarti}). 

\begin{figure}[ht]
\vspace{-1.5cm}
\hspace{-2.2cm}
\centerline{\psfig{figure=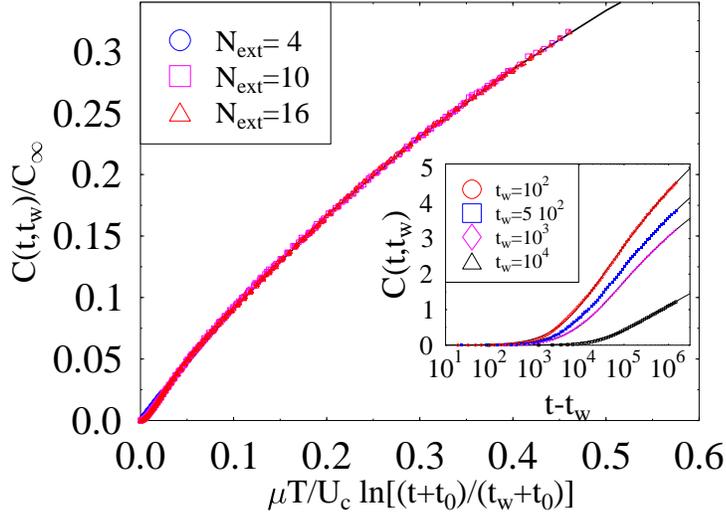,width=8cm,angle=-90}}
\vspace{-2.0cm}
\caption{{\bf Inset}
Logarithmic time relaxation of the two-times correlation function, $C(t,t_w)$, 
as a function of $t-t_w$ for the shown values of $t_w$. 
Data are recorded  at $T=0.1$ (i.e., below $T_g$) and $N_{ext}=16$. 
The continuous lines are logarithmic fits from eq.(\ref{log_fits}).
{\bf Main Frame} Off equilibrium dynamical scaling.
The relaxation data of $C(t,t_w)$ from the inset and those recorded at 
$N_{ext}=4,10,16$ for each of the shown $t_w$ are superimposed on the
same master function. The asymptotic scaling is $C(t,t_w)\sim{\cal S}(t/t_w)$.
}
\label{rel_sca}
\end{figure}          

\begin{figure}[ht]
\vspace{-1.5cm}
\hspace{-1.8cm}
\centerline{\psfig{figure=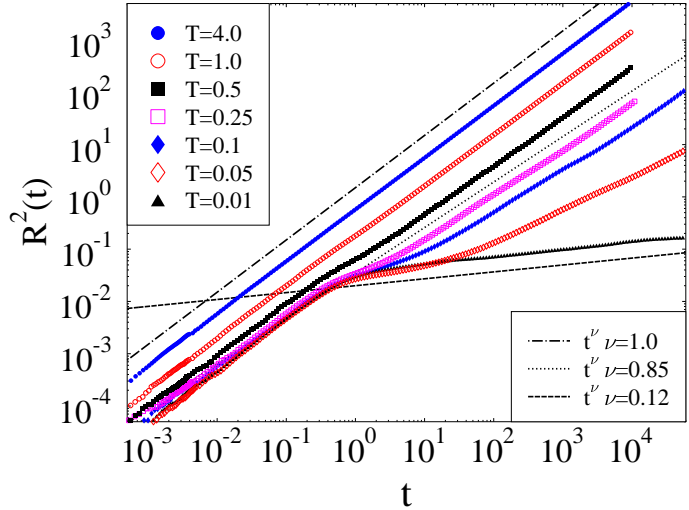,width=8cm,angle=-90}}
\vspace{-2.0cm}
\caption{The vortex mean square displacement $R^2(t)$
at $N_{ext}=10$ for several temperatures.
Below $T_g\sim 0.25$, $R^2(t)$ is strongly subdiffusive: $R^2(t)\sim t^{\nu}$
with $\nu<1$. Straight lines are guides for the eye.
}
\label{r2_fig}
\end{figure}      

Since below $T_g$ relaxation times are huge, one might expect that the motion 
of the particles essentially stops. Instead, as shown below,
the off equilibrium dynamics has remarkably rich ``aging'' properties. 
In the inset of Fig.\ref{rel_sca} we show that
$C(t,t_w)$, at $T=0.1$, exhibits strong ``aging'': 
$C$ depends on both times $t$ {\em and} $t_w$; in particular, 
its evolution (see Fig.\ref{rel_sca}) is slower the older is the ``age'' 
$t_w$ (``stiffening''). 
In the entire low $T$ region ($T<T_g$), after a short initial power law
behaviour, $C(t,t_w)$ can be well
fitted by a generalisation of a known interpolation formula,
often experimentally used \cite{blatter,yeshurun}, which now depends on
the {\em waiting time}, $t_w$:
\begin{equation}
C(t,t_w)\simeq C_{\infty}
\left\{1-\left[1+
\frac{\mu T}{U_c}\ln\left(\frac{t+t_0}{t_w+t_0}\right)\right]^{-1/\mu}\right\}
\label{log_fits}
\end{equation} 

Eq.(\ref{log_fits}) (in agreement with the general scenario of 
Ref.\cite{CoNi}) implies the presence of {\em scaling properties} 
of purely dynamical origin (see Fig.\ref{rel_sca}): 
for times large enough (but smaller than the
equilibration time), $C$ is a universal function of the ratio $t/t_w$:
$C(t,t_w)\sim {\cal S}(t/t_w)$. 

In experiments about vortex creep in superconductors a crossover is 
usually found from a low $T$ region with logarithmic creep to an high 
$T$ region with typically power law or stretched exponential 
relaxations (see for instance \cite{niderost} and Ref.s in \cite{yeshurun}). 
In particular, aging in magnetic creep has been recently observed in BSCCO  
samples \cite{Papad}. 
We also recall that the above phenomena are intriguingly common to many 
different systems ranging from polymers, to supercooled liquids \cite{Angell}, 
spin glasses \cite{glass_rev,DFNC_iflg}, granular media \cite{NC_gm}.

\subsection{Vortex mean square displacement}

The microscopic origin of the above features in the system dynamics can be 
understood by considering the vortex mean square displacement, 
$R^2(t)$ (plotted in Fig.\ref{r2_fig} for $N_{ext}=10$). At high enough $T$,
$R^2(t)$ is linear in $t$ (in agreement with experiments and MD simulations, 
see \cite{Kes_nature,Monier} and ref.s therein), 
but at lower temperatures it shows a pronounced bending. Finally,
below $T_g$, the process becomes strongly subdiffusive:
\begin{equation}
R^2(t)\sim t^{\nu}
\end{equation}
with $\nu<< 1$. From this point of view, $T_g$ is the location of a sort of 
structural arrest of the system, where particles displacement is dramatically 
suppressed. 
Each vortex is caged by other neighbouring vortices for long times. 
The system dynamics needs large scale ``cooperative rearrangements'' to 
relax \cite{Angell}. 
Interestingly, a very similar scenario has been recorded in real
superconducting samples (see for instance \cite{Fuchs}).

\begin{figure}[ht]
\vspace{-1.5cm}
\hspace{-2.2cm}
\centerline{\psfig{figure=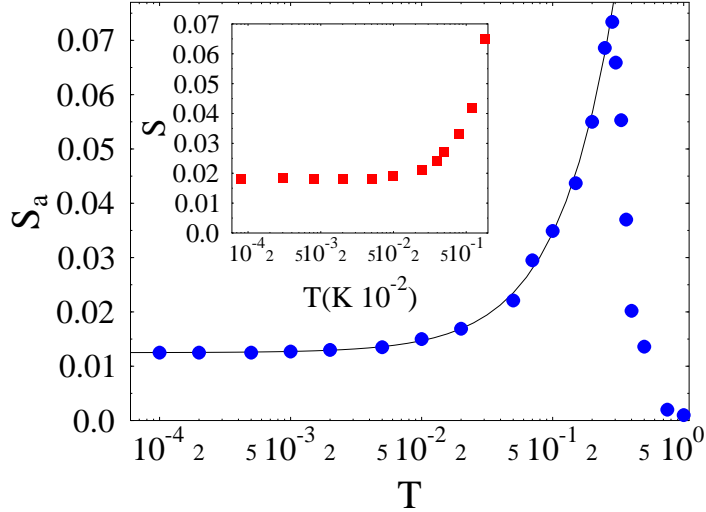,width=8cm,angle=-90}}
\vspace{-2.0cm}
\caption{{\bf Main frame} The creep rate, $S_a$, in the ROM model for 
$N_{ext}=10$ as a function of the temperature, $T$, in units of $A_0$ 
($\kappa^*=0.28$, $\gamma=10^{-3}$). The superimposed line is a linear fit.
{\bf Inset} Creep rate, $S$, in a BSCCO single crystal at 880 Oe 
(from Aupke {\em et al.} [53]).} 
\label{S_T}
\end{figure}

\subsection{The creep rate, $S_a$, for $T\rightarrow 0$}

With the insight on the system dynamics obtained in the previous 
sections, we can now understand an other intriguing experimental observation 
\cite{mota-fruchter-stein,aupke,mota_org,mota_hf} about vortex matter:
even at very low temperatures (where activated processes should be absent) 
magnetic relaxation does take place. 
This surprising phenomenon, previously interpreted 
in terms of ``quantum tunnelling'' of vortices \cite{blatter}, 
is also found in the present purely ``classical'' vortex model.
More generally, we show here that a non-zero creep rate for $T\rightarrow 0$ 
is to be expected in systems ``aging'' in their off equilibrium dynamics. 

\begin{figure}[ht]
\vspace{-1.6cm}
\hspace{-1.8cm}
\centerline{\psfig{figure=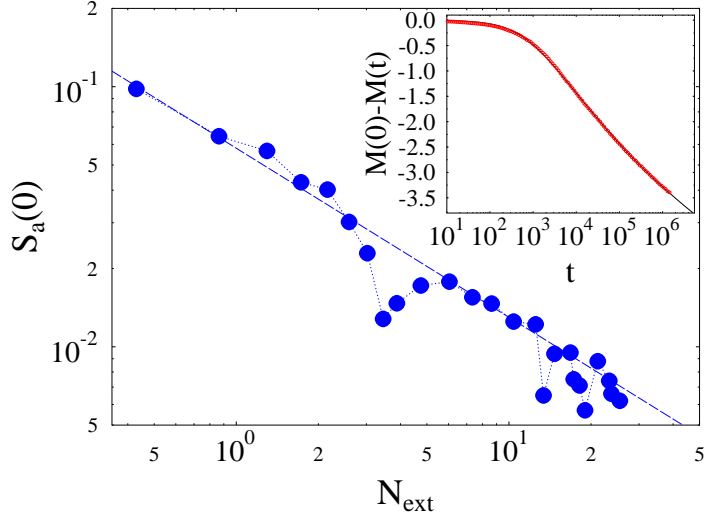,width=8cm,angle=-90}}
\vspace{-2.0cm}
\caption{{\bf Main frame} The $T\rightarrow 0$ limit of the creep rate, $S_a$, 
in the ROM model as a function of the applied field $N_{ext}$ 
(for $T=10^{-4}$ and $\gamma=10^{-3}$). 
The superimposed dashed curve is a power law to guide the eye.  
{\bf Inset: } 
The relaxation of the magnetisation, $M(t)$, in the model 
for $N_{ext}=10$ and $T=0.25$ as a function of time ($\gamma=10^{-3}$). 
The continuous line is the logarithmic fit of the text.} 
\label{S_0}
\end{figure}

Experiments investigate the temperature dependence of the creep rate, $S_a$, 
(see Fig. \ref{S_T}), where 
\begin{equation}
S_a=\left|{\partial \ln (M)\over \partial \ln(t)}\right|
\end{equation} 
($S_a$ is, as usual, averaged in some given temporal window 
\cite{yeshurun,mota-fruchter-stein,aupke,mota_org,mota_hf}). 
When the temperature is extremely low the magnetisation still logarithmically 
relaxes (see inset Fig.\ref{S_0}), and in both experiments and in our 
simulations, $S_a$ approaches a {\em finite} plateau, $S_a(0)>0$, 
for $T\rightarrow 0$. In Fig.\ref{S_T}, we plot the creep rate, $S_a$,
as a function of $T$. For comparison we present experimental data in BSCCO 
(from Ref.\cite{aupke}) as inset (note that the values of $S_a$ in our model 
and in real samples are very similar). In particular, we find that a linear 
fit of $S_a(T)$ in the low $T$ regime is very satisfactory (see Fig.\ref{S_T}):
\begin{equation}
S_a(T)=S_a^0+\sigma T
\end{equation}
where both $S_a^0$ and $\sigma$ are functions of the applied field
$N_{ext}$. In the present model, as much as in experiments 
\cite{mota_org,yeshurun}, $S_a(T)$ is non monotonous in $T$: at high $T$ 
it starts decreasing 
(this is due to the fact that, for a given observation window, 
at higher $T$ the system gets closer to equilibrium, see 
Fig.\ref{CT1par_varie_temp} and \cite{nj_1}). 

By varying the applied field 
we find a range of values for $S_a^0$ very similar to experimental ones 
\cite{mota-fruchter-stein,aupke,mota_org,mota_hf}
(see Fig.\ref{S_0}). In particular, 
$S_a^0$ seems to decrease on average by increasing the field $N_{ext}$. 
The overall behaviour can be roughly interpolated with a power law:
$S_a^0(N_{ext})\simeq (N_{ext}/N_0)^{-x}$, where, for $\kappa^*=0.28$, 
$N_0\simeq 0.01$ and $x\simeq 0.6$. As shown in Fig.\ref{S_0}, 
the presence of a small exponent, $x$, implies that 
sensible variations in $S_0^a$ can be seen only 
by changing $N_{ext}$ of orders of magnitude. 
Note that in Fig. \ref{S_0} the dips in the $S_a(0)$ versus $N_{ext}$ data 
found at certain values of $N_{ext}$ (namely around 3, 13, and 20) 
are statistically significant. They are located respectively close to
the region of the 
low field order-disorder transition (see Fig.\ref{mfpd}), the 
2nd peak transition and the reentrant high field order-disorder transition. 

In the slow off equilibrium relaxation at very low temperatures no activation
over barriers occurs and the system simply wanders in its very high
dimensional phase space through the few channels where no energy increase
is required. We have already shown that at very low $T$, the system 
equilibration time, $\tau(T)$, diverges exponentially. 
In that region, the typical observation time windows, $t_{obs}$, are
such that $t_{obs}/\tau\ll 1$, and the system is in the early stage of its 
off equilibrium relaxation from its initial state. This is schematically the 
origin of the flattening of $S_a$ at very low $T$ \cite{nj_1}.
Notice that, in a system observed for an exponentially long time,
i.e., for $t_{obs}/\tau\gg 1$, the creep rate, $S_a$, would indeed go to zero.

\begin{figure}[ht]
\vspace{-1.5cm}
\centerline{
\hspace{-2.5cm}\psfig{figure=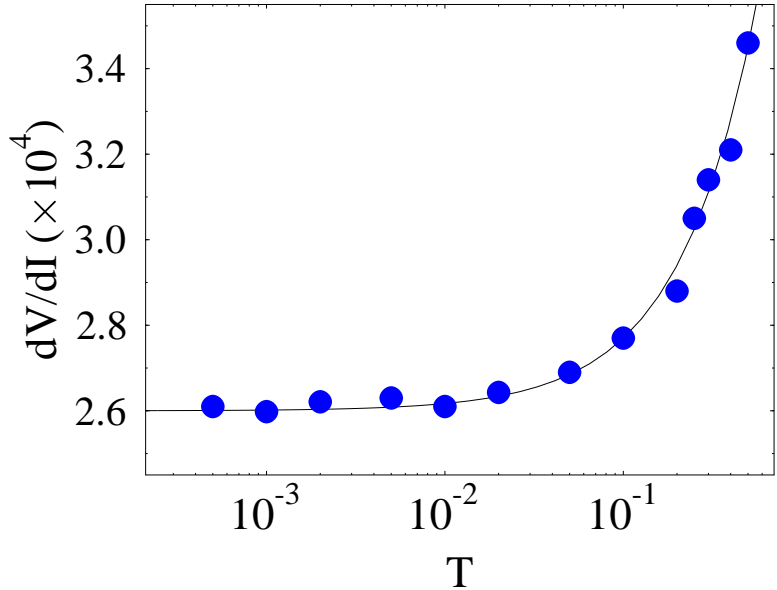,width=9cm,angle=-90}}
\vspace{-1.6cm}
\caption{The differential resistivity, $\rho=dV/dI$, in the ROM model is 
plotted as a function of the temperature, $T$, for $N_{ext}=10$. 
The continuous superimposed curve is a linear fit.
The saturation of $\rho(T)$ for $T\rightarrow 0$ well compares with 
the one of the creep rate, $S(T)$. 
}
\label{fig14}
\end{figure}

Interestingly, our model along with a saturation of the creep rate, $S(T)$, 
also shows a saturation of the dissipation in the limit $T\rightarrow 0$. 
We show in Fig.\ref{fig14} the differential resistivity, 
$\rho(T)=dV/dI$, measured for the same value of the model parameters 
used in the calculation of the creep rate $S(T)$ in Fig.\ref{S_T} 
(the precise definition of $V$ and $I$ is postponed to the next section 
where we consider the I-V characteristics). 
The continuous curve superimposed to $\rho(T)$ in Fig.\ref{fig14} 
corresponds to the linear fit $\rho(T)=\rho_0+\sigma_{\rho} T$. 
These results clearly show a saturation in $\rho(T)$ at low $T$ towards 
a finite value, in a way similar to the one recorded in $S(T)$.

The present scenario, where off equilibrium phenomena dominate the anomalous 
low $T$ creep, is supported by the experimental discovery of ``aging'' in the 
relaxation \cite{Bhatta,Giller,Roy,Paltiel,Papad,higgins,andrei}. 
In fact, strong discrepancies are found 
between ``quantum creep theory'' predictions \cite{blatter} 
and the observed low $T$ relaxation in many compounds 
\cite{nj_1,mota_hf,hoekstra}. 
Interestingly, a unified picture begins 
to emerge of magnetic and transport properties. 
This will be discussed further in the next section. 

\section{The I-V characteristic}

Vortex flow in driven type II superconductors also shows strong memory and 
history dependent effects. Here, we outline the relations with magnetic 
properties and propose a scenario for a broad set
of these kind of phenomena ranging from {\em ``rejuvenation''} and
{\em ``stiffening''} of the system response, to {\em ``memory''}
and {\em ``irreversibility''} in I-V characteristics.
In relation to recent experimental results \cite{higgins,andrei}, we discuss
in particular the nature of ``memory'' effects observed in the response of
the system to an external drive, i.e., the I-V characteristic.
Our model explains the peculiar form of such a ``memory'' of vortex flow
at finite $T$ and other ``anomalous'' properties such as the time dependence 
of critical currents. The essential step is, again, 
to identify the relevant time scales in the dynamics.

The system is zero field cooled and prepared by increasing $N_{ext}$
at constant rate, $\gamma$, up to the working value (here, $N_{ext}=10$).
Then we monitor the system relaxation after applying a
drive, $I$ (due to an external current which induce a Lorentz force on 
vortices), in the $y$-direction.
As in similar driven lattice gases \cite{KLS}, the effect of the drive
is simulated by introducing a bias in the Metropolis coupling of the
system to the thermal bath: a particle can jump to a neighbouring
site with a probability min$\{1,\exp[-(\Delta{\cal H}-\epsilon I)/T]\}$.
Here, $\Delta{\cal H}$ is the change in ${\cal H}$ after the jump
and $\epsilon = \pm 1$ for a particle trying to hop along or opposite
to the direction of the drive and $\epsilon = 0$ if orthogonal jumps occur. 
A drive $I$ generates a voltage $V$ \cite{hyman}:
\begin{equation}
V(t)=\langle v_a(t) \rangle
\end{equation}
where $v_a(t)={\overline v(t)}$ is an average vortex ``velocity'' at time $t$
\cite{nj_1}. 
Here, 
$v(t)= {1\over L} \sum_i v_i(t)$ is the instantaneous flow ``velocity'',
$v_i(t)=\pm 1$ if the vortex $i$ at time $t$ moves along or opposite
to the direction of the drive $I$ and $v_i=0$ otherwise.

\begin{figure}[ht]
\vspace{-1.6cm}
\hspace{-1.8cm}
\centerline{\psfig{figure=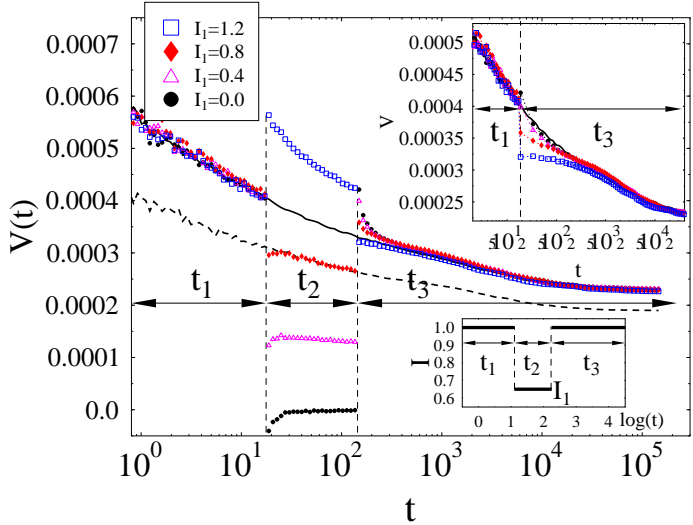,width=8cm,angle=-90}}
\vspace{-2.0cm}
\caption{In the ROM model the voltage, $V(t)$, is plotted as a function of 
time at $T=0.1$ for a drive $I=1$. As shown in the lower inset,
after a time lag $t_1$, the drive is abruptly
changed to $I_1$ for a time $t_2$ and finally it is set back to its
previous value. When $I$ is switched to $I_1$ the system seems to
{\em ``rejuvenate''}: it suddenly restarts its relaxation along the path
it would have had if $I=I_1$ at all times (consider the
continuous and dashed bold curves, corresponding to $I=I_1=1$ and
$I=I_1=0.8$, plotted for comparison).
By restoring $I$ after $t_2$, the system shows a strong form of ``memory'':
if $t_2$ and $I_1$ are small enough (see text) the relaxation of
$V(t)$ restarts where it was at $t_1$. However, if $t_2$ and $I_1$ are
too large, this is not the case, as shown in the upper inset.
In this sense, the above is an {\em ``imperfect memory''}.}
\label{4}
\end{figure}

\subsection{Memory effects in driven vortex flow}

We analyse 
a striking manifestation of ``memory'' observed in experiments where the
drive is cyclically changed in the low $T$ region \cite{andrei}. 
A drive $I$ is applied to the system
and, after a time $t_1$, abruptly changed to a new value $I_1$; finally,
after waiting a time $t_2$, the previous $I$ is restored and the system
evolves for a further $t_3$ (see lower inset of Fig.\ref{4}).
The measured $V(t)$ is shown in the main panel of Fig.\ref{4} for $T=0.1$.
A first observation is that after the switch to $I_1$ the system
seems to abruptly reinitiate its relaxation approximately as if it has
always been at $I_1$ (see for example the dashed curve in Fig.\ref{4}),
a phenomenon known as {\em ``rejuvenation''}
in thermal cycling of spin-glasses and other glassy systems
\cite{bouchaud}. The more surprising fact is, however, that for $I_1$ small
enough (say $I_1\ll I^*$, $I^*$ to be quantitatively defined below)
when the value $I$ of the drive is restored the voltage
relaxation seems to restart from where it was at $t_1$, i.e., where it stopped
before the switch to $I_1$ (see Fig.\ref{4}). Actually, if one ``cuts''
the evolution during $t_2$ and ``glues'' together those during $t_1$ and
$t_3$, an {\em almost} perfect matching is observed (see upper inset of
Fig.\ref{4}). What is happening during $t_2$ is that the system is trapped
in some metastable states, but {\em not} completely frozen as shown by a small
magnetic, as well as voltage, relaxation. 
These non trivial ``memory'' effects are experimentally found in vortex matter
\cite{andrei} and glassy systems \cite{bouchaud}. We call them
a form of {\em ``imperfect memory''}, because they tend to disappear when the
time spent at $I_1$ becomes too long or, equivalently
(as explained below) when, for a given $t_2$, $I_1$ becomes too high,
as shown in the upper inset of Fig.\ref{4}. 

\begin{figure}[ht]
\vspace{-1.6cm}
\hspace{-1.8cm}
\centerline{\psfig{figure=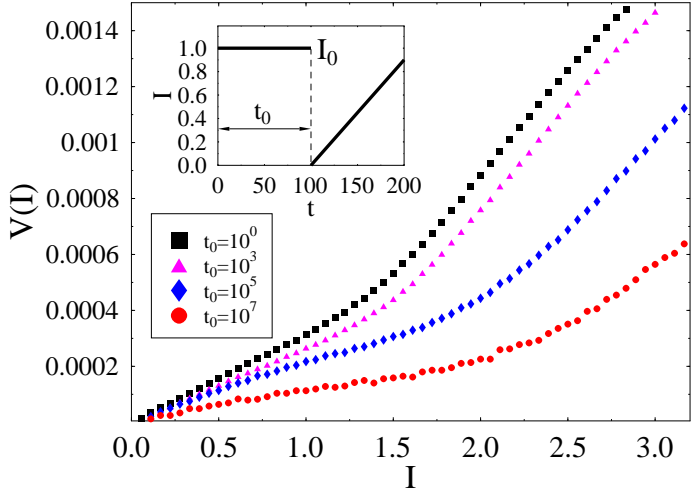,width=8cm,angle=-90}}
\vspace{-2.0cm}
\caption{The I-V obtained at $T=0.1$ by ramping $I$ after keeping the system
in presence of a drive $I_0=1$ for a time $t_0$ as shown in the inset.
The response, $V$, is ``aging'' (i.e., depends on $t_0$) and,
more specifically, {\em stiffening}: it is smaller the longer $t_0$.}
\label{2}
\end{figure}

\subsection{History dependent I-V}

We now turn to the time 
dependent properties of the current-voltage characteristic.
As in real experiments on vortex matter \cite{andrei},
we let the system undergo a current step of hight $I_0$ for a time $t_0$
before starting to record the I-V by ramping $I$, as sketched in the inset of
Fig.\ref{2}. Fig.\ref{2} shows (for $T=0.1$) that the I-V depends on the
waiting time $t_0$. The system response is ``aging'': the longer
$t_0$ the smaller the response, a phenomenon known as {\em ``stiffening''}
in glass formers \cite{Angell,glass_rev,bouchaud}. 
These effects are manifested in the violation of time translation invariance 
of two times correlation functions, already discussed. 

These simulations also reproduce the experimentally found time dependence of
the critical current \cite{andrei}. Usually, one defines an effective critical
current, $I_c^{eff}$, as the point where $V$ becomes larger than a given
threshold (say $V_{thr}=10^{-5}$ in our case): one then finds that
$I_c^{eff}$ is $t_0$ and $I_0$ dependent (like in experiments \cite{andrei}
$I_c^{eff}$ is slowly increasing with $t_0$, see Fig.\ref{2}).  

\begin{figure}[ht]
\vspace{-1.6cm}
\hspace{-2cm}
\centerline{\psfig{figure=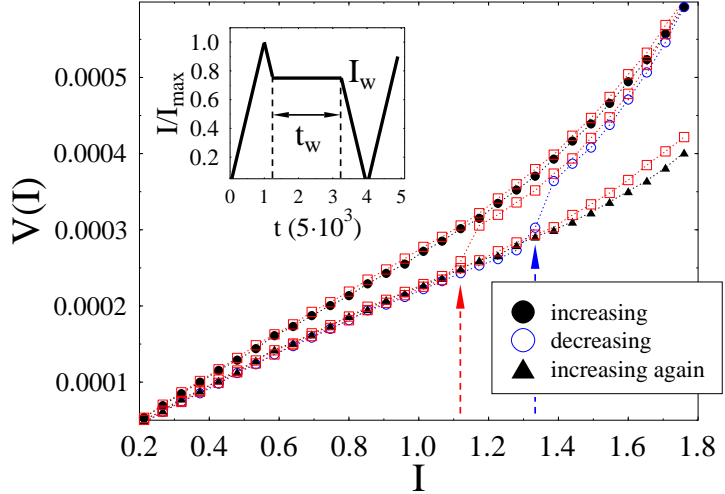,width=8cm,angle=-90}}
\vspace{-1.9cm}
\caption{The I-V is measured at $T=0.1$ during cycles of $I$ 
(see also the inset): 
$I$ is at first increased up to $I_{max}$ (filled circles); along the 
descending branch of the cycle (empty circles), when $I=I_w$ (in the main 
panel the $I_w$'s, for two cycles, are located by the arrows) the drive is 
kept fixed for a time 
$t_w=10^4$ and then the cycle restarted; finally, $I$ is ramped up again 
(filled triangles). 
For $I>I_w$, the first increasing ramp and the decreasing one (resp. filled 
and empty circles) do not completely match, showing {\em irreversibility} in 
the I-V. After waiting $t_w$ at $I_w$, a much larger separation is seen. 
However, by raising $I$ again (filled triangles) a strong {\em memory} 
is observed: the system doesn't follow the first branch (filled circles), 
but the decreasing one (empty circles). 
Furthermore, in a cycle with a lower $I_w$ (squares), 
the {\em same} branches are found.} 
\label{5}
\end{figure}

It is interesting to consider another current cycling experiment which 
outlines the concurrent presence of irreversibility and memory effects 
(see Fig.\ref{5}). The I-V is measured by ramping $I$ up to some value 
$I_{max}$. Then $I$ is ramped back to zero, but at a given value $I_{w}$ 
the system is let to evolve 
for a long time $t_w$. Finally, $I$ is ramped up again (see inset of 
Fig.\ref{5}). The resulting {\em irreversible} $V(I)$ 
is shown in Fig.\ref{5}. 
For $I>I_{w}$ the decreasing branch of the plot (empty circles) 
slightly deviates from the increasing one (filled circles), showing the 
appearance of {\em irreversibility}. This is even more apparent after $t_w$: 
for $I<I_{w}$ the two paths are clearly different. Interestingly, upon 
increasing $I$ again (filled triangles), 
$V(I)$ doesn't match the first increasing branch, but the latest, the 
decreasing one: in this sense there is 
coexistence of memory and irreversibility.
Also very interesting is that by repeating the cycle with a new $I_{w}$ 
(squares), the system approximately follows the {\em same} branches. 
This non-reversible behaviour is also found in other glassy systems 
\cite{bouchaud}. However, spin glasses, for instance, seem to show the 
presence of the so called {\em chaos effects} \cite{bouchaud,chaos_eff}. 
The chaos effect is absent in our system as it is also in other ordinary 
glass formers \cite{glass_rev,bouchaud}. 
This kind of interplay between irreversibility and memory can be checked
experimentally in superconductors and thereby assess the present scenario.  

\subsection{Differential resistivity}

\begin{figure}[ht]
\vspace{-1.6cm}
\hspace{-1.8cm}
\centerline{\psfig{figure=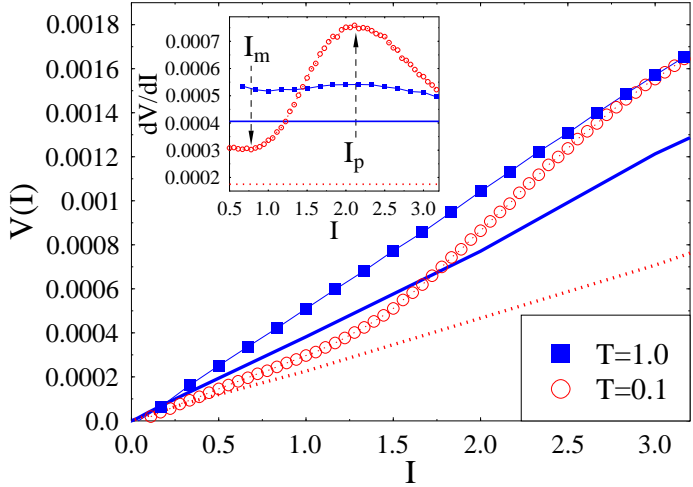,width=8cm,angle=-90}}
\vspace{-2.0cm}
\caption{
The I-V is recorded by ramping $I$ for the shown $T$. 
The continuous and dotted curves (resp. $T=1, 0.1$) are the 
{\em asymptotic} I-V, i.e., those where, for a given $I$, $V$ is measured 
after waiting $t=1.5\cdot 10^5$ (see Fig.\ref{1}).
{\bf Inset} The differential resistivity, $R=dV/dI$, for the same data 
of the main panel. The horizontal lines are from a linear fit to the 
{\em asymptotic} I-V. 
The characteristic values $I_m$ and $I_p$ roughly locate crossover points in 
the ``short time'' $R$, which, however, disappear if $t\rightarrow\infty$.} 
\label{3}
\end{figure}

In Fig.\ref{3}, we plot the I-V recorded after ramping $I$ 
at $T=1$ (filled squares) and $T=0.1$ (open circles). 
The low $T$ I-V has the typical S shaped form experimentally found 
\cite{yeshurun,cohen,kwok,safar,higgins,andrei}, 
but, since we are above $T_c$, 
this is only an effect of short times of observation. 
The linear continuous functions in Fig.\ref{3} 
are, in fact, the {\em asymptotic I-V}, i.e., those recorded after applying a 
drive $I$ and measuring $V$ in the long time regime (for $t=1.5\cdot 10^5$ 
in Fig.\ref{1}). 
The same analysis applies to the differential resistivity, $R=dV/dI$, shown 
in the inset of Fig.\ref{3}. Also here one might think to see some 
characteristic regimes 
(defined, for instance, by the values $I_m$, $I_p$ of the inset of Fig.\ref{3})
in the ``short time'' $R(I)$. 
They might be the off stationarity, finite temperature 
rests of crossovers between different plastic channels flow regimes typically 
found at $T=0$, as discussed in \cite{henrik,Nori,bassler,Monier} 
and references therein (see also \cite{eqtr}). Here, the 
linear behaviour of the asymptotic I-V indeed shows that the crossovers 
in the ``short time'' $R(I)$ tend to slowly disappear with time, thus 
they cannot correspond to transitions among different driven stationary phases 
\cite{eqtr,henrik,Nori,Gronbech-Jensen,Zimanyi,higgins,andrei}. 
This conclusion holds despite the regular behaviour of $I_m$ and $I_p$ 
with $T$ also experimentally seen (for instance $I_p$ seems to rapidly grow 
with $T$). 
An intrinsic structure in $R$ can possibly be observed at sufficiently lower 
currents and temperatures \cite{eqtr}. 

\subsection{Voltage relaxation}

\begin{figure}[ht]
\vspace{-1.6cm}
\hspace{-1.8cm}
\centerline{\psfig{figure=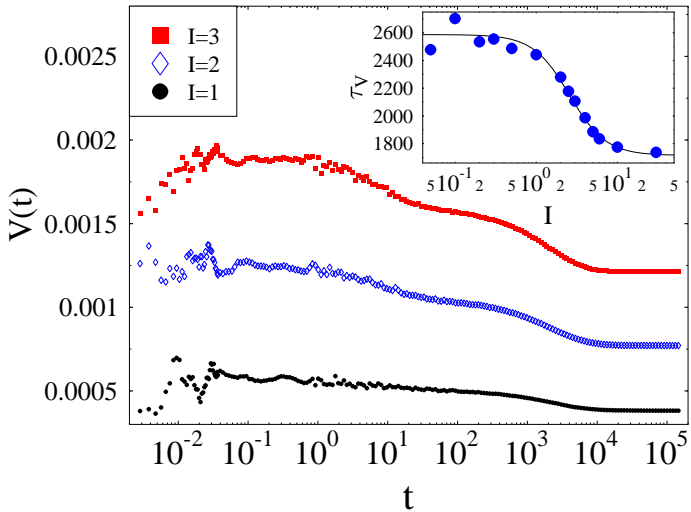,width=8cm,angle=-90}}
\vspace{-2.0cm}
\caption{The time evolution of the response function, $V(t)$, for the shown
values of the drive $I$ (at $T=1$ and $N_{ext}=10$).
In the asymptotic regime $V(t)$ is well
fitted with: $V(t)\propto \exp[-(t/\tau_V)^{\beta}]$.
{\bf Inset} The characteristic scale of relaxation, $\tau_V(I)$, as a function
of $I$. For $I\rightarrow 0$, $\tau_V(I)$ seems to
saturate to a {\em finite} value which implies $I_c=0$.}
\label{1}
\end{figure}

The natural step to understand the above observations is the 
identification of the characteristic time scales of the driven dynamics, 
which in the present model can be well accomplished. This we now discuss. 
Upon applying a small drive, $I$, the system response, $V$, relaxes
following a pattern with two very different parts:
at first a rapidly changing non-linear response is seen,
later followed by a very slow decrease towards stationarity
(see $V(t)$ in Fig.\ref{1} for $T=1$ and $I\in\{1,2,3\}$).
For instance, for $I=3$ in a time interval $\Delta t\simeq 2\cdot 10^{-1}$,
$V$ leaps from about zero to $\Delta V_i\sim 2\cdot 10^{-3}$, corresponding
to a rate $r_i=|\Delta V_i/\Delta t|\sim 10^{-2}$. This is to be compared with
the rate of the subsequent slow relaxation from, say, $t=2\cdot 10^{-1}$
to $t=10^4$, $r_f \sim 10^{-7}$:
$r_i$ and $r_f$ differ of 5 orders of magnitude.

In agreement with experimental findings \cite{higgins,andrei,danna},
the slow relaxation of $V(t)$ has a characteristic double step
structure, which asymptotically can be well fitted by stretched exponentials
\cite{nota_log}: $V(t)\propto \exp(-t/\tau_V)^{\beta}$.
The above long time fit defines the characteristic asymptotic scale, $\tau_V$,
of relaxation. The exponent $\beta$ and $\tau_V$
are a function of $I$, $T$ and $N_{ext}$ (see inset Fig.\ref{1}):
in particular $\tau_V(I)$ decreases with $I$ and seems to approach
a {\em finite plateau} for $I<I^*$, with $I^*\simeq O(1)$.
In this sense, the presence of a drive $I$ makes the approach
to stationarity faster and has an effect similar to an increase in $T$. 

The outlined properties of $\tau_V$ clearly explain the history
dependent effects in the experiments previously considered.
For instance, the ``imperfect memory'', discussed in Fig.\ref{4}, is caused by
the presence of a long, but finite, scale $\tau_V$ in the problem:
for a given $I_1$ the system seems to be
frozen whenever observed on times scales smaller than $\tau_V(I_1)$.
Thus, if $t_2$ is short enough ($t_2<\tau_V(I_1)$) the system preserves a
strong ``memory'' of its state at $t_1$.
The weakening of such a ``memory'' found for higher currents $I_1$
in Fig.\ref{4}, is also a consequence of the strong decrease of
$\tau_V(I)$ with $I$. The phenomenon of ``rejuvenation'' (see Fig.\ref{4})
is, in turn, a consequence of the presence of the extremely fast first part
of relaxation found in $V(t)$ upon applying a drive
and of the above long term memory.
The existence of the slow part in the $V(t)$ relaxation
also affects the ``stiffening'' of the response in the I-V of Fig.\ref{2},
which is due to the non-stationarity of the vortex flow
on scales smaller that $\tau_V$.
Actually, in Fig.\ref{2}, for a given $I$ the value of $V$ on the different
curves corresponds to the system being probed at different stages of its
non-stationary evolution. 
Finally, in brief, 
the fact that $\tau_V(I)$ is smaller at high currents, $I$, and 
larger at small $I$ (and $T$), is responsible for the 
surprisingly concomitant effects of irreversibility and memory of Fig.\ref{5}. 

The origin of these time dependent properties of the driven flow,
and in turn those of I-V's, traces back to the
concurrent vortex creep and reorganisation of vortex domains. In fact,
both with or without an external drive, the
system evolves in presence of a Bean like profile 
(see inset of Fig.\ref{mfpd}) which in turn relaxes. 
An important discovery is that the characteristic times scales 
of voltage and magnetic relaxation are approximately proportional \cite{nj_1}.
This outlines that the  non-stationary voltage relaxation is
structurally related to the reorganisation of vortices during the
creep (a fact confirmed by recent experiments \cite{Paltiel}).  

\section{Conclusions}

In conclusion, we showed that the replica mean field theory and Monte Carlo 
simulations of a schematic statistical mechanics lattice model\cite{nj_1} for 
vortices in type-II superconductors (a system of particles diffusing in a 
pinning landscape) offer a comprehensive framework of off equilibrium 
magnetic and transport properties observed in vortex matter. 
Off equilibrium phenomena in many respect are known to show strong 
``universalities'' \cite{Angell,glass_rev}. 
In fact, here we considered either a mean field or a two dimensional version 
of the ROM model, which, interestingly, reproduces a very broad spectrum 
of experimental results. Molecular Dynamics simulations of more 
realistic systems, when existing, seem to confirm the present scenario 
\cite{nj_1,henrik,Nori,Gronbech-Jensen,Zimanyi}, and, even if very demanding 
in the low $T$ and high fields region, they can be an essential test for it. 

We have seen that the model shows a reentrant phase diagram in the 
field-temperature plane (B,T), analogous to what observed in vortex matter. 
More specifically, we discussed the off equilibrium, ``aging'', properties of 
magnetic creep. 
At low temperatures a crossover point is found, $T_g(N_{ext})$, 
where the system relaxation times become exponentially large. 
They seem to diverge \`a la Vogel-Tamman-Fulcher at a lower
temperature, $T_c(N_{ext})$, where an ``ideal'' glassy transition point can be
located. Magnetic creep changes its structure around $T_g$: 
above $T_g$ it shows power laws asymptotically followed by 
stretched exponential saturation; below $T_g$ it is logarithmic.
This corresponds to a change in microscopic vortex motion: from diffusive 
(above $T_g$) to strongly subdiffusive \cite{nj_1}. 
We showed that in the low temperature region the system is
very far from equilibrium and its time correlation functions, no longer 
invariant under time translations, have interesting 
dynamical scaling properties analogous to those of other ``aging" systems. 
The above ``off equilibrium'' scenario also explains the surprising 
experimental discovery of a finite creep rate, $S_a>0$, when $T\rightarrow 0$ 
(previously interpreted in terms of ``quantum tunnelling''
of vortices \cite{blatter}) in our purely ``classical'' model. 

At not too high temperatures (but still well above $T_g$), magnetisation loops 
are typically found when $M$ is plotted as a function of the applied field, 
including a definite ``second peak" when the 
Ginzburg-Landau parameter is not too low. The ``second peak" is associated 
with a new phase transition in the system. This can be difficult
to see in experiments because samples can be significantly out
of equilibrium, as shown by the dependences of the loops on the external
field sweep rate. 

Vortex flow in driven type II superconductors also shows strong memory and 
history dependent effects. We have shown how creep and transport properties 
in driven media are related. We proposed a scenario for a broad set
of these kind of phenomena ranging from ``rejuvenation'' and
``stiffening'' of the system response, ``memory''
and ``irreversibility'' in I-V characteristics, to history dependent 
critical currents. 

The emerging unifying scenario of magnetic and transport properties in 
vortex physics has interesting relations with off equilibrium phenomena 
in other glass formers and complex fluids such as random magnets and 
supercooled liquids.

{\bf Acknowledgement} HJJ is supported by the British EPSRC.
MN acknowledges support from INFM-PRA(HOP) and INFM-PCI.


\begin{thebibliography}{9}
\bibitem{nota_notc} Notice that in several cases the ``ideal'' transition is 
found at zero temperature, but the systems have apparent ``glassy'' dynamics 
(see examples in \cite{glass_rev}).
\bibitem{nota_l0} Note that for $\l_o\rightarrow \xi$ one recovers the 
original system dealt with in MD.
\bibitem{nota_L} Here 
$A_0^p=0.3 A_0$, $N_{c2}=27$, $p=1/2$ and, if not differently stated, 
$A_1=0.28 A_0$ (the ${\cal H}$ parameters can be related to the real material 
parameters \cite{nj_1}). $L=32$, but the results are checked up to $L=128$.
\bibitem{nota_MC} Since the general dynamical features of $C$ and $M$ are 
very similar, for clarity here we mainly discuss $C$.
\bibitem{nota_log} At lower $T$ inverse logarithmic relaxations are
found \cite{nj_1}.
\end{thebibliography}
\end{document}